# Role of heterogeneity in dictating tumorigenesis in epithelial tissues

Sindhu M.[1] and Medhavi Vishwakarma[1]*

1. Center for BioSystems Science and Engineering, Indian Institute of Science, Bangalore, 560012

Correspondence: medhavi@iisc.ac.in

**Abstract:**

Biological systems across various length and time scales are noisy, including tissues. Why are biological tissues inherently chaotic? Does heterogeneity play a role in determining the physiology and pathology of tissues? How do physical and biochemical heterogeneity crosstalk to dictate tissue function? In this review, we begin with a brief primer on heterogeneity in biological tissues. Then, we take examples from recent literature indicating functional relevance of biochemical and physical heterogeneity and discuss the impact of heterogeneity on tissue function and pathology. We take specific examples from studies on epithelial tissues to discuss the potential role of inherent tissue heterogeneity in tumorigenesis.

**Introduction:**

Noise serves functional roles in biology across different scales, contrary to intuition. At the micro-scale, biased Brownian motion of the Myosin head over the Actin filament results in muscle contraction[1,2]. At the macro scale, intrinsic variations in velocity and direction among individual fishes bring about collectivity in fish schooling[3]. At the intermediate mesoscale heterogeneity in the position and alignment of ciliated cells in the lung tissue facilitates optimal ciliary flow clearance[4]. Biological tissues display heterogeneity in terms of biochemical and mechanical processes within the cells[5]. Part of this heterogeneity is deterministic, such as variations arising from cellular differentiation into distinct cell types[6–9]. However, a lot of biological heterogeneity is observed as stochastic noise[10–14]. While the role of deterministic heterogeneity in tissues is intuitive, stochastic heterogeneity is often characterized as noise[15] and the physiological relevance of this noise remain largely elusive. We explore the relevance of stochastic heterogeneity in epithelial function in the following section.

**Stochastic heterogeneity, not just noise:**

Stochastic heterogeneity within biological tissues represents the inherent randomness in the biochemical processes within the cells. Various factors, such as asynchronized cell cycles[16,17], differential metabolic[18,19] & epigenetic states[20,21], and the asymmetric distribution of organelles[17,22], contribute to this randomness. Multiple studies have characterized the biological noise using single-cell measurements[20,23,24] such as flow cytometry[25,26], fluorescence microscopy[27,28], real-time PCR[29] and microfluidics[30]. Additionally, cellular variations in the gene expression over time[31–34] due to transcriptional bursts[35–37] also contributes to stochastic heterogeneity. In addition to this biochemical heterogeneity, physical heterogeneity has been reported in tissues of epithelial origin, initially revealed by spatiotemporal variations in cell-cell and cell-substrate forces across epithelium[38](Fig 1B.).

Several studies point to possible functional roles for stochastic heterogeneity in epithelia. For instance, the patchy arrangement of cilia in the mouse airway epithelium generates a locally heterogeneous flow of airway clearance that is globally efficient[39]. Heterogeneity and stochastic growth have also been proposed to regulate biliary epithelial tissue remodelling, which is central to liver regeneration[11]. Further, T cells can infiltrate through the endothelial Basement membrane containing Laminin 4 but not through endothelial BM containing Laminin 5[40] during Experimental Autoimmune Encephalomyelitis (EAE) (Fig.1c). Heterogeneity is also associated with pathological conditions such as ageing[41,42]. For instance, cell-cell variations in levels of protein expression is higher in old mice as compared to young mice[43]. Additionally, stochastic physical heterogeneity is shown to regulate essential physiological functions of the tissues. For instance, mechanical heterogeneity regulates collective cell migration during epithelial wound closure[44,45]. Another evidence was reported in developing zebrafish where fluctuations in cellular stresses regulates anteroposterior body axis formation [46,47]. Studies also implicate physical heterogeneity in pathological situations such as Asthma[48], pulmonary fibrosis[49] and cancer metastasis[50,51].

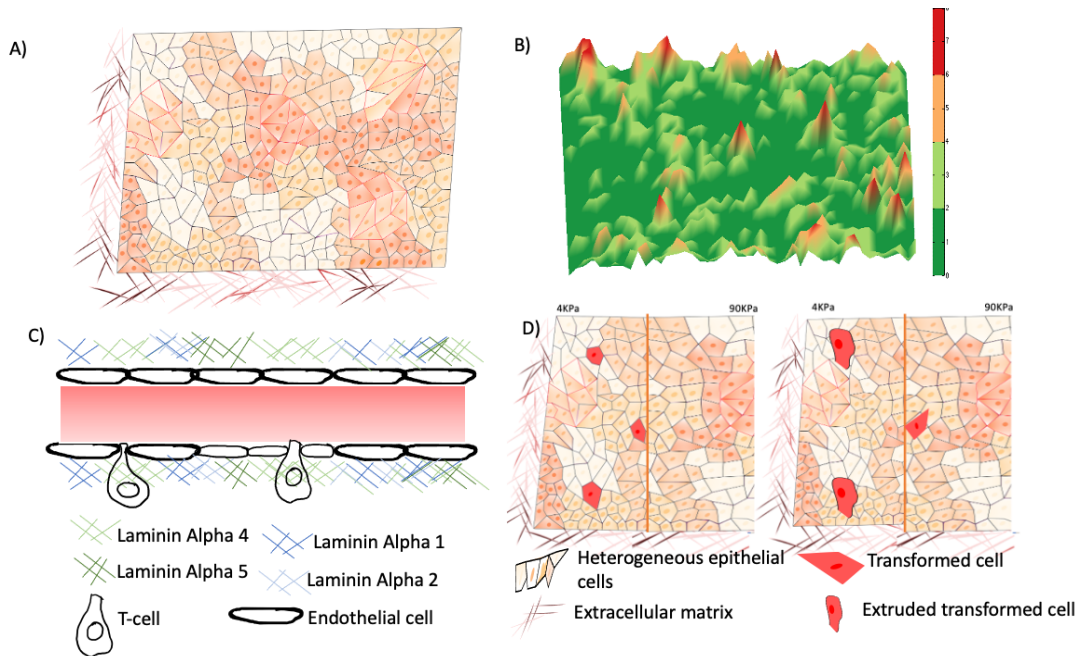

Fig 1: Relevance of epithelial heterogeneity in tissue function- A) Cartoon representation of a heterogeneous epithelial monolayer, B) Rugged landscape of stress distribution from Monolayer Stress Microscopy, C) T-cell extravasation depends on the Laminin heterogeneity in the basement membrane, D) Mutant cells at the interface of soft and stiff substrate migrate to stiff substrate and escape extrusion.

Overall, these studies indicate functional relevance of heterogeneity in biological tissues. The relevance of inherent heterogeneity on disease initiation remains largely unexplored. In the next section, we summarize reports suggesting that inherent tissue heterogeneity might regulate both the initiation and progression of diseases, emphasizing on cancer initiation.

**Heterogeneity and Cancer initiation:**
Cancer initiates within the habitat of the tissue and cancer cells grow in the space of host cells[52]. Lately, it has been recognized that cancer cells enter into a survival battle with the surrounding host cells during tumorigenesis, and can kill the host cells to make space for their growth[52–57]. Excitingly, in many cases, similar competitive interactions allow host cells to recognize the emerging cancer cells and extrude them out of the tissue[58–64]. An understanding on how cancer cells outcompete the surrounding host cells during cancer initiation may open doors to novel targeted therapies[52]. Studies in the past few years have demonstrated clonal competition and selection during tumorigenesis in both *in-vitro*[65–67] and *in-vivo*[52–57,66–77] model systems. In addition, signalling molecules playing mechanistic roles in cellular-competition during tumorigenesis have been described in several studies[78–87]. However, description of cell competition remains incomplete without factoring in the inherent cellular heterogeneity within the tissues. How does tissue heterogeneity affect selection of mutants during cancer initiation and progression? Do biochemical and physical heterogeneity within the tissues affect the competitive interactions within cancer cells and host cells? Slaughter's concept of field cancerization[88] suggest that biochemical heterogeneity caused by the genetic and epigenetic differences within the cells would play a crucial role in determining growth or suppression of cancer cells. To this end, a recent study demonstrate that genetic heterogeneity across the tissue plays an essential role in determining survival of early neoplasms in mouse esophageal tissue[89]. Another study in mice thyroid tissue shows that intrinsic properties of thyroid follicles determined fate of mutant cells. Follicular heterogeneity and thyroid tissue organization dictated the fate of BRAF mutant cells[90] with an increased propensity of BRAF mutants to develop tumor in the postnatal Thyroid.

Besides the genetic and mutational landscape of tissues, studies suggest that tumour initiation may also depend on the mechanical landscape of tissues. While mutant HRas-V12 cells are successfully extruded out from the epithelial monolayer cultured on soft substrates, extrusion of mutants is impaired on stiffer substrates. Furthermore, mutant cells at the interface of soft and stiff substrates migrate to the stiff substrate by durotaxis and evade extrusion[84], suggesting that the ECM stiffness landscape may profoundly affect tumorigenesis. In another study, in the pancreatic tissue, tumorigenesis was found to depend on the tissue architecture. It was observed that in small pancreatic ducts, tumor growth was away from the duct, whereas in large ducts tumor growth was inward towards the lumen, suggesting that tension imbalance and tissue curvature may play a crucial role in epithelial tumorigenesis[91]. Another study show that in stratified epithelia of skin tissue, tumor progression in the early stages is shaped by forces exerted as a result of tissue structure[92]. Cells with HRas mutation in mouse embryonic skin produce rigid cells with high Keratin levels which are unable to dissipate compressive forces, and hence rupture the basement membrane and invade the underlying tissue[93].

To develop a framework explaining these experimental observations describing relevance of heterogeneity and stochasticity in cancer initiation, application of concepts from physics such as non-linear dynamics and critical transitions might be helpful, which will be the focus of the next section.

**The physics of heterogeneity and cancer initiation.**

Physicists and mathematicians view cells as high dimensional complex systems[94,95] with thousands of genes involved in gene regulatory networks. According to the framework of high dimensional complex systems, interactions of gene regulatory networks in cells could result in multiple possible states. Out of the multiple possible configurations, the stable states are called 'Attractors', and multipotent cells differentiate into one of these stable states[95–97]. Cancer may be viewed as the change in cell state from a normal state to an aberrant state[98]. Stochastic variations may provide cells with the impetus needed to overcome the potential barrier required for a state-change. For instance, heterogeneity is high at tissue boundaries, and tissues boundaries are also tumour hotspots in Squamous cell carcinoma[61].

In the framework of chaotic systems, cancer progression is seen as a change in the type of attractor, from the Torus attractor (see box 1) in premalignant lesions to the strange attractor (See box 1) in advanced stages of cancer, resulting in chaos[98,99] (see box). The change in the type of attractor is accompanied by period-doubling(see box) and sequential bifurcations(see box), and cells are able to adopt multiple aberrant metastable states as a result, which is proposed as

> **A GLOSSARY ON CHAOS:**
>
> **Chaos:** Chaos is a property of non-linear systems which adopt multiple unpredictable states with no discernible pattern and the system fluctuates in seemingly random ways.
>
> **Period doubling:** The state of non-linear dynamic systems may be predicted before the onset of chaos. The onset of chaos is preceded by period-doubling and sequential bifurcations. Bifurcations are said to occur when the system proceeds from one initial steady state to two alternating steady states and subsequently to four and so on. Since the number of states keep doubling, the frequency of transitions between the states is also doubled, known as period doubling.
>
> **Chaos onset:** As the system proceeds through sequential bifurcations and period doubling, after a certain critical point, the system adopts multiple states without any predictable patterns, and this critical point represents the onset of chaos.
>
> **Critical transitions**: The transition due to these bifurcations is seen as an abrupt change in the state of the complex system, called a 'critical transition'. Critical transitions are drastic changes in the state of the system despite no apparent change in the driver. Critical transitions are different from stochastic transitions which occur far from tipping point due to noise.
>
> **Early Warning Signals(EWS):** Critical transitions are preceded by characteristic signatures in the time series data. An increase in the Autocorrelation function, variability, and skewness just before the tipping point may be used to predict an impending critical transition and is knows as Early Warning Signals.
>
> **Sensitive dependence on initial conditions:** In non-linear dynamic systems, it is possible to predict future states when we know the exact initial conditions and it is deterministic in principle. However, even miniscule changes in the initial conditions leads to vastly different predicted states, known as 'Sensitive dependence on initial conditions', a hallmark of chaos.
>
> **Attractors:** Despite the unpredictability in chaotic systems, a collection of initial conditions converges to specific final states called attractors. There may be different types of attractors such as the well-known butterfly shaped Lorenz attractor, Torus attractor and the strange attractor, depending upon the equations dictating the chaotic systems.

*Box 1: Glossary explaining the common vocabulary associated with nonlinear dynamic systems and chaotic theory*

an alternative explanation for Intra tumor heterogeneity[98,99]. The transition due to these bifurcations is seen as an abrupt change in the state of the complex system, which may be predicted by Early Warning Signals(EWS). EWS (see box 1)have been detected in cell-fate determination of multipotent blood progenitor cells[96], and studies have suggested the use EWS for the early detection of colorectal Cancer[100] and breast cancer[101].

Another characteristic of complex non-linear dynamic systems is self-repeating fractal patterns. They are fractional dimensional structures that appear similar at different magnification levels and are a common theme in nature, observed in snowflakes, coastlines, and branching neurons. Fractal patterns have also been reported in tumour growth patterns, and studies have used fractal dimension as an indicator of tumour malignancy and carcinogenesis[98,102,103].

**Conclusions**

While genetic and molecular etiology of cells and the changes in the composition of surrounding extracellular environment have been explored as therapeutic targets to treat many diseases originating from epithelial tissues, including cancer[104,105]; how diseased conditions arise and how noise within the tissue may regulate initiation of pathology remains elusive. Understanding the underlying heterogeneity in tissues and its role in tumorigenesis will help us appreciate the organizational principles and emergent self-regulatory properties of collective living systems. Acknowledging the role of heterogeneity in tissues might prompt biologists to move away from time and population averaged studies, and lead to important discoveries. While the impact of mechanical and biochemical heterogeneity on tissue function have been studied in isolation, the exciting possibility of their interdependency being an inherent tissue property and contributing to emergent tissue functions has not been explored. Extrapolating lessons from epithelia and pondering on the relevance of noise in biological systems across multiple scales raises an intriguing question- could scale-invariant fundamental principles rooted in noise contribute to the emergent self-regulatory properties unique to life?